\documentclass[12pt,showpacs]{revtex4}
\usepackage[english]{babel}
\usepackage{graphicx}
\usepackage{epsfig}

\begin{document}
\title{Reply to ``Comment on\\
`The polarizability of the pion: no conflict\\
between dispersion theory and chiral perturbation theory' ''} \normalsize
\author{B. Pasquini}
\affiliation{
Dipartimento di Fisica Nucleare e Teorica, Universit\`{a} degli Studi di Pavia, and\\
Istituto Nazionale di Fisica Nucleare, Sezione di Pavia, I-27100 Pavia, Italy}
\author{D. Drechsel and S. Scherer}
\affiliation{ Institut f\"ur Kernphysik, Johannes Gutenberg Universit\"at,
D-55099 Mainz, Germany}
\begin{abstract}
We reply to the Comment by  Fil'kov and Kashevarov, arXiv:0805.4486 [hep-ph].
We show that the discrepancies between ChPT and dispersion theory, 
reported for the polarizability of the pion, result
from applying dispersion theory to non-analytic functions.
\end{abstract}
\date{\today}
\maketitle
The electric ($\alpha$) and magnetic ($\beta$) polarizabilities of a composite system such as the pion are
elementary structure constants, just as its size and shape.
They can be studied by applying electromagnetic fields to the
system, that is, by the Compton
scattering process $\gamma + \pi \rightarrow \gamma + \pi$ or the
crossed-channel reaction $\gamma + \gamma \rightarrow \pi + \pi$.
Within the framework of the partially conserved axial-vector (PCAC)
hypothesis and current algebra, the
polarizabilities of the charged pion were related to the
radiative decay $\pi^+\to e^+\nu_e\gamma$~\cite{Terentev:1972ix}.
Chiral perturbation theory (ChPT) at
leading non-trivial order, ${\cal O}(p^4)$,
confirmed this result, $\alpha_{\pi^+}=-\beta_{\pi^+}\sim \bar l_\Delta$~\cite{Bijnens:1987dc},
where $\bar l_\Delta\equiv(\bar l_6-\bar l_5)$ is a linear
combination of scale-independent parameters of the Gasser and
Leutwyler Lagrangian~\cite{Gasser:1983yg}. At ${\cal O}(p^4)$
this combination is related to the ratio of the
pion axial-vector form factor $F_A$ and the vector form factor
$F_V$ of radiative pion beta decay, $F_A/F_V={\bar{l}}_\Delta/6$ ~\cite{Gasser:1983yg}.
Once this ratio is known, chiral symmetry makes an {\emph {absolute}}
prediction at ${\cal O}(p^4)$, $\alpha_{\pi^+}=2.64\pm 0.09$, here and
in the following in units of $10^{-4}\, \mbox{fm}^3$.
Corrections to this leading-order result were
calculated at ${\cal O}(p^6)$ and turned out to be rather small~\cite{Burgi:1996qi,Gasser:2006qa}.
In particular, these corrections were estimated through vector-meson saturation of low-energy constants.
The largest correction was found to stem from the $\omega$, which contributed about $-0.7$ to
$(\alpha-\beta)_{\pi^0}$~\cite{Donoghue:1993kw,Bellucci:1994eb},
whereas the $\rho$ and other vector mesons yielded much smaller contributions. These findings are at variance
with the results of Fil'kov and Kashevarov who obtained an $\omega$ contribution of $-12.56$ to
$(\alpha-\beta)_{\pi^0}$~\cite{Fil'kov:2005ss} and similarly increased effects from other vector mesons.
On condition that the ChPT estimates for the low-energy constants are correct,
the following predictions for the polarizabilities provide a significant
test of ChPT~\cite{Gasser:2006qa}:
\begin{equation}
(\alpha + \beta)_{\pi^+} = 0.16 \pm 0.1\,, \quad (\alpha - \beta)_{\pi^+} = 5.7 \pm 1.0\,.
\label{eq:1.1}\\
\end{equation}
The results of ChPT are in sharp contrast with predictions based on
dispersion relations~\cite{Fil'kov:2005ss},
\begin{equation}
(\alpha + \beta)_{\pi^+} = 0.17 \pm 0.02\,, \quad(\alpha - \beta)_{\pi^+} = 13.60 \pm 2.15 \,.
\label{eq:1.3}
\end{equation}
We attribute  this discrepancy to unphysical singularities introduced by Ref.~\cite{Fil'kov:2005ss}
in a region very close to the Compton threshold, the
point at which the polarizabilities are determined.\\
The model of Ref.~\cite{Fil'kov:2005ss} describes the contribution of a vector meson
$V=\{\rho, \omega\}$ to Compton scattering by an energy-dependent
coupling constant $g(s)$ at the vertex $\gamma + \pi \rightarrow V$
and a vector meson propagator $1/\{(M-i\Gamma (s)/2)^2-s\}$, with $M$ the mass of
the vector meson and $\Gamma (s)$ its energy-dependent width. The term
quadratic in $\Gamma (s)^2$ is neglected (small-width approximation).
The vector meson contributions to the amplitudes take the form
\begin{equation}
\label{eq:3.1}
M^{+-}(s)=\frac{4\, g(s)^2}{M^2-s-iM\Gamma (s)}\,,
\quad  M^{++}(s)= -s \, M^{+-}(s)\,,
\end{equation}
where $M\Gamma (s)$ and  $g(s)$ correspond to $\Gamma_0$ and $g_V^2$
as defined by Eq.~(3) of Ref.~\cite{Fil'kov:2009}.
The vector meson contributions to the polarizabilities are derived
from the amplitudes by
\begin{equation}
\label{eq:3.4}
\alpha + \beta=\frac {m} {2 \pi} M^{+-}(s=m^2),\quad
\alpha - \beta= \frac {1} {2 \pi m} M^{++}(s=m^2)\,.
\end{equation}
Combining these equations with Eq.~(\ref{eq:3.1}), we obtain
\begin{equation}
\label{eq:3.5}
\alpha+\beta= -(\alpha- \beta) \quad \Rightarrow \quad \alpha= 0\,.
\end{equation}
The (quasi-static) electromagnetic transition from the pion ($J^P=0^-$)
to the intermediate vector meson ($J^P=1^-$) is a magnetic
dipole transition yielding a paramagnetic contribution to $\beta$ and, as a consequence,
a ratio $R=(\alpha- \beta)/(\alpha+\beta)=-1$. To the contrary,
Fil'kov and Kashevarov predict a ratio $R\approx-20$ and a large electric
polarizability $\alpha$ (see Table~\ref{tab:1}). Even more surprising,
the latter carries a negative sign, a result only
possible in a relativistic quantum field theory. In conclusion, the results of Ref.~\cite{Fil'kov:2005ss} are
at variance with Eq.~(\ref{eq:3.5}). Because the relations $M^{++}(s)= -s \, M^{+-}(s)$ and
$\alpha= 0$ follow directly from evaluation of the Feynman diagram for
$\gamma + \pi (J^P=0^-) \rightarrow \rho/\omega (J^P=1^-) \rightarrow \gamma + \pi (J^P=0^-)$,
Table~\ref{tab:1} shows that the treatment of vector mesons in Ref.~\cite{Fil'kov:2005ss} can not
be understood in terms of a diagrammatic approach.\\
\begin{table}
\begin{center}
\begin{tabular}{|l|cc|cc|}
\hline
meson & $\alpha+\beta$ & $ \alpha - \beta$ & $\alpha$ & $\beta$ \\
\hline\hline
$\rho $ & $0.063$ & $-1.15$  & $ -0.54$ & $0.61$ \\
$\omega$ & $0.721$ & $-12.56$ & $- 5.92$ & $6.64$ \\
\hline
\end{tabular}
\end{center}
\caption{\label{tab:1}Results of Ref.~\cite{Fil'kov:2005ss}
for the contributions of $\rho$ and $\omega$ mesons to the
polarizabilities of charged and neutral pions, respectively, in units of $10^{-4}\, {\rm {fm}}^3$.}
\end{table}
The apparent discrepancy between the two approaches can be traced to the specific forms
for the imaginary part of the Compton amplitudes~\cite{Fil'kov:2005ss}, which serve as input
to determine the polarizabilities at the Compton threshold ($s=m^2,\, t=0$) by dispersion integrals. In order
to obtain amplitudes with good properties at high energies,
Fil'kov and Kashevarov introduce energy-dependent widths describing the correct threshold behavior of the
two-pion intermediate states as well as energy-dependent coupling constants with square-root singularities, e.g., $g(s)^2 \sim 1/\sqrt{s}$
in Eq.~(3) of Ref.~\cite{Fil'kov:2009}.
The resulting amplitudes fulfill the conditions to set up dispersion relations:\\
(i) The amplitudes are analytic on the physical Riemann sheet except for {\emph {isolated}} points on the real
axis. As an example, the $s$-channel singularities of Ref.~\cite{Fil'kov:2005ss} are situated (i) at the threshold for two-pion states ($s=4m^2$), which
leads to the physical cut, and (ii) at the origin of the Mandelstam plane ($s=0$), which leads to
an unphysical cut.\\
(ii) The amplitudes are square integrable along any line parallel to the real axis, albeit at the expense of an unphysical
cut due to the square-root singularity at $s=0$.\\
Up to this point, the assumptions allow one to apply Titchmarsh's theorem: the real and the imaginary parts of the amplitudes are
Hilbert transforms, that is, they are related by dispersion relations.
However, Fil'kov and Kashevarov drop the contribution from the unphysical cut by setting
the imaginary part equal to zero 
``below the threshold of two-pion production''. 
In whatever manner one implements their statement 
``Im{$M^{++}(s,t)=0$ for $s<4m^2$''~\cite{Fil'kov:2009} in the complex $s$ plane, the procedure will inevitably introduce a ``wall''
in the complex plane, separating a region with finite values of the imaginary part from zero values. Because all the points on this wall
become non-analytic ones, the function itself is no longer analytic (note: condition (i) allows only for {\emph {isolated}} singularities).
Furthermore, in the region of vanishing imaginary parts, the only possible analytic function is a real constant everywhere. And unless
this constant vanishes, one easily finds that also condition (ii) breaks down. These arguments show the inconsistency of first
introducing the factor $1/\sqrt{s}$ for convergence at large energies and later ignoring its consequence, the unphysical cut,
at low energies. The results of this inconsistency are: (i) the vector meson effects are grossly overestimated and
(ii) the magnetic dipole transition $\gamma + \pi \rightarrow \rho/\omega$ yields a {\emph {large negative}} contribution
to the electric polarizability. In their Comment~\cite{Fil'kov:2009}, Fil'kov and Kashevarov motivate the neglect of the left-hand singularities
by Regge model and bootstrap conceptions, such that the unphysical singularities in the
$s$ channel are canceled by similar singularities in the $t$ channel.
This argument may be disputed even for the evaluation of $\alpha - \beta$ by means of dispersion relations at $u=m^2$,
because it is difficult to understand how $s$ and $t$ channel singularities, lying at different points on the integration path, should
exactly cancel. The argument is even more questionable for the combination
$\alpha + \beta$, which is obtained by forward dispersion relations ($t=0$) in Ref.~\cite{Fil'kov:2005ss}.
In this case both $s$ and $u$ channel singularities contribute, and because
the amplitude must be symmetric under $s \leftrightarrow u$ crossing, the postulated cancelation is clearly impossible.
Similar problems show up for the exchange of other mesons~\cite{Pasquini:2008ep}. In particular,
the  $1/\sqrt{t}$ factor for $\sigma$ exchange in the $t$ channel leads to a diverging amplitude $M^{++}(t)$ at $t=0$,
the point at which the polarizability is to be predicted. Even apart from dispersion relations, we would not recommend to
fit the data in the measurable region by functions that approach infinity at or near the point to which one wants to extrapolate. In conclusion,
the reported discrepancies between ChPT and dispersion theory result from applying the latter theory to non-analytic
functions.
\end{document}